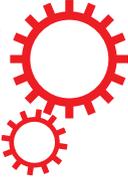

# Higher ventricular rate during atrial fibrillation relates to increased cerebral hypoperfusions and hypertensive events

Andrea Saglietto[1], Stefania Scarsoglio[2], Luca Ridolfi[3], Fiorenzo Gaita[1,4] & Matteo Anselmino[1]

Atrial fibrillation (AF) is associated with cognitive impairment/dementia, independently of clinical cerebrovascular events (stroke/TIA). One of the plausible mechanisms is the occurrence of AF-induced transient critical hemodynamic events; however, it is presently unknown, if ventricular response rate during AF may impact on cerebral hemodynamics. AF was simulated at different ventricular rates (50, 70, 90, 110, 130 bpm) by two coupled lumped parameter validated models (systemic and cerebral circulation), and compared to corresponding control normal sinus rhythm simulations (NSR). Hemodynamic outcomes and occurrence of critical events (hypoperfusions and hypertensive events) were assessed along the internal carotid artery-middle cerebral artery pathway up to the capillary-venous bed. At the distal cerebral circle level (downstream middle cerebral artery), increasing ventricular rates lead to a reduced heart rate-related dampening of hemodynamic signals compared to NSR ($p=0.003$ and $0.002$ for flow rate and pressure, respectively). This response causes a significant progressive increase in critical events in the distal cerebral circle ($p < 0.001$) as ventricular rate increases during AF. On the other side, at the lowest ventricular response rates (HR 50 bpm), at the systemic-proximal cerebral circle level (up to middle cerebral artery) hypoperfusions ($p < 0.001$) occur more commonly, compared to faster AF simulations. This computational study suggests that higher ventricular rates relate to a progressive increase in critical cerebral hemodynamic events (hypoperfusions and hypertensive events) at the distal cerebral circle. Thus, a rate control strategy aiming to around 60 bpm could be beneficial in terms on cognitive outcomes in patients with permanent AF.

During the last two decades, atrial fibrillation (AF), the most common cardiac tachyarrhythmia, has become one of the most relevant public health problems[1]. This growing epidemiological burden warrants answers to currently pending questions regarding AF. In particular, it has emerged that AF is associated with an increased risk of dementia and cognitive impairment, even in anticoagulated patients[2] in the absence of clinical strokes[3,4]. Several mechanisms have been proposed to clarify this association[5–7], such as silent cerebral ischemia (SCI), microbleeds, altered cerebral blood flow dynamics and pro-inflammatory conditions.

Among these possible contributors, the hypothesis of an altered cerebral blood flow dynamics during AF has been the least investigated, most likely due to the evident concerns related to a direct sampling in the cerebral circulatory system. In addition, the currently adopted non-invasive techniques in the field of cerebral hemodynamics, as transcranial doppler (TCD)[8] ultrasonography, lack the resolving power to provide insights, in terms of flow and pressure signals, on the portion downstream the three cerebral arteries (anterior, middle and posterior). Given the paucity of clinical data, our group recently run a validated computational model to simulate cerebral hemodynamics during AF[9,10], concluding that AF is associated, when compared to normal sinus rhythm (NSR), with an altered cerebral hemodynamics, characterized by transient hypoperfusions and hypertensive events in the deep cerebral circle.

[1]Division of Cardiology, "Città della Salute e della Scienza di Torino" Hospital, Department of Medical Sciences, University of Turin, Turin, Italy. [2]Department of Mechanical and Aerospace Engineering, Politecnico di Torino, Torino, Italy. [3]Department of Environmental, Land and Infrastructure Engineering, Politecnico di Torino, Torino, Italy. [4]Cardiology Department, Clinica Pinna Pintor, Turin, Italy. Andrea Saglietto and Stefania Scarsoglio contributed equally. Correspondence and requests for materials should be addressed to S.S. (email: stefania.scarsoglio@polito.it)

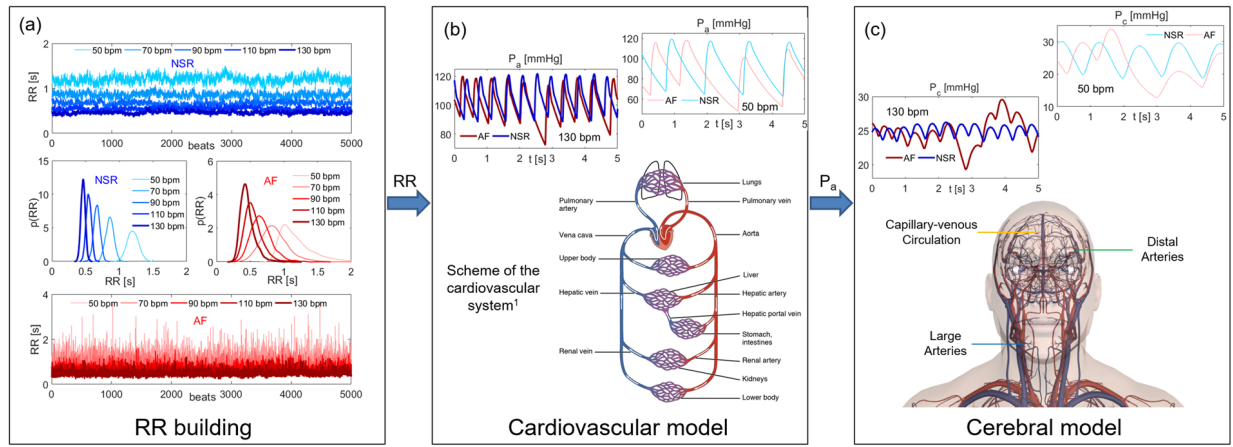

**Figure 1.** Scheme of the computational algorithm. (**a**) RR Building. 5000 RR beats extracted in NSR and AF at the selected HR (50, 70, 90, 110, 130 bpm) and the corresponding probability distribution functions. Blue: NSR. Red: AF. (**b**) Cardiovascular model. Schematic representation of the cardiovascular system, together with examples of the resulting $P_a$ time series in NSR and AF conditions (HR = 50 and 130 bpm). [1]Figure obtained from https://commons.wikimedia.org/wiki/File:2101_Blood_Flow_Through_the_Heart.jpg under Creative Commons Attribution 3.0 Unported license (https://creativecommons.org/licenses/by/3.0/deed.en). (**c**) Cerebral model. Schematic representation of the cerebral model evidencing the three main regions (large arteries, distal arteries, capillary-venous circulation), together with examples of the resulting $P_c$ time series in NSR and AF conditions (HR = 50 and 130 bpm). $P_a$ and $P_c$ time series are taken within the same temporal interval.

If the hemodynamic consequences of AF, therefore, candidate *per se* as a contributing factor to the non-embolic cerebral events and cognitive decline related to the arrhythmia, another open clinical question refers to the ideal heart rate target to achieve during rate control strategy of patients with permanent AF. In general, clinical evidence on this topic is scarce. The RACE II clinical trial[11], albeit with widely discussed limitations[12], suggested that lenient (resting heart rate <110 beats per minute) and strict (resting heart rate <80 beats per minute and heart rate during moderate exercise <110 beats per minute) rate control strategies did not differ in terms of mid-term cardiovascular outcomes. However, cerebral hemodynamics, cognitive impairment and dementia were not within the studied outcomes. Interestingly, Cacciatore *et al.*[13] demonstrated, on a small group of AF patients with cognitive impairment, that low (<50 bpm) and high (>90 bpm) mean ventricular responses related to increased progression towards dementia.

Given these presumptions, the present study aims to investigate, based on a validated computational model, if the cerebral hemodynamic alterations induced by AF are modulated by mean ventricular response.

## Methods

**Computational algorithm.** The stochastic modelling of the AF cerebral hemodynamics has been recently proposed[9,10,14] and relies on a three-steps algorithm, which is sketched in Fig. 1. The algorithm combines a stochastic extraction of the heart beating with a sequence of two lumped parameter models. The systemic arterial pressure, $P_a$, output of the cardiovascular model is exploited as the forcing input of the subsequent cerebral model.

*(a) Beating features: NSR and AF at different HR.* The heart beating, RR [s], is defined as the temporal interval between two consecutive heart beats, while the heart rate HR [bpm] is the number of heart beats per minute. We used, both in NSR and AF conditions, artificially-built RR intervals to span the range of average HR between 50 and 130 bpm, thus avoiding the patient-specific characteristics inherited by real RR beating.

Normal RR heart beats are extracted from a Gaussian distribution, which is the typical distribution observed during sinus rhythm for RR. Since normal heart beating is correlated and represents an example of pink noise[15], RR extraction is carried out according to the pink noise temporal structure[16,17]. Standard deviation values, $\sigma$, are determined considering that the coefficients of determination, cv, similarly range in the same interval [0.05, 0.14] for all the HR considered[18]. Thus, cv was kept constant and equal to 0.07 from 50 to 130 bpm.

AF distribution, instead, is for 60–65% of the cases unimodal[15,17] and is described by the superposition of two statistically independent times, $RR = \varphi + \eta$. $\varphi$ is taken from a Gaussian distribution and the extraction is based on the correlated pink noise. $\eta$ is instead drawn from an exponential distribution (with rate parameter $\gamma$) and the beating extraction relies on the uncorrelated white noise. The resulting AF beatings are thus represented by an exponentially modified Gaussian (EMG) distribution. The standard deviation values, $\sigma$, are determined keeping the coefficient of variation, cv, constant at each HR and equal to 0.24, as recommended by Tateno *et al.*[19]. The rate parameter, $\gamma$, instead is a linear function of the mean RR ($\gamma = -9.2RR + 14.6$), as proposed in a previous work[20].

Since the RR intervals are based on beating features[16] and have been validated and tested over clinically measured data[15,17,21], they were adopted as the most suitable and reliable RR time-series to model NSR and AF conditions. By this approach, the resulting RR extraction intrinsically contains the chronotropic effects due to the heart rate regulation, which differently act during NSR and AF.

To guarantee the statistical stationarity of the results, 5000 cardiac cycles are extracted for each configuration. The 5000 RR beats extracted in NSR and AF conditions together with the corresponding RR probability distribution functions are displayed in Fig. 1a, while Supplementary Table 1 in the Supplementary Information summarizes the main statistics of the beating RR during NSR and AF.

*(b) Cardiovascular model.* Following RR extraction, the cardiovascular model was run to obtain systemic arterial pressure ($P_a$). The model, proposed by Korakianitis and Shi[22], was validated during AF in resting conditions over more than 30 clinical datasets[16,23], and then exploited to study the cardiovascular response in different conditions and pathologies related to AF[20,24,25]. Through a network of compliances, resistances and inductances, the cardiovascular dynamics includes the systemic and venous circuits together with an active representation of the four cardiac chambers, and is expressed in terms of pressures, flow rates, volumes and valve opening angles. Both atria are imposed as passive to simulate AF conditions, while they can actively contract during NSR.

In addition, a baroreceptor model[26] was coupled to the proposed cardiovascular model. The short-term baroregulation accounts for the inotropic effect of both ventricles, as well as the control of the systemic vasculature (peripheral arterial resistances, unstressed volume of the venous system, and venous compliance). As mentioned above, the chronotropic effects due to the heart rate regulation are instead implicitly taken into account by the RR extraction. Details of the governing equations and model parameters are offered in the Supplementary Information.

By solving the cardiovascular model, the resulting systemic arterial pressures, $P_a$, was then used as forcing inputs for the forthcoming cerebral model. A schematic representation of the complete cardiovascular system is reported in Fig. 1b, along with examples of $P_a$ time series in NSR and AF at different HRs.

*(c) Cerebral model.* The cerebral model is based on a lumped parameterization of the arterial and venous cerebral circulation, along with the cerebrovascular control mechanisms of autoregulation and $CO_2$ reactivity[27]. The model is able to reproduce several different pathological conditions characterized by heterogeneity in cerebrovascular hemodynamics and has been validated in normal conditions up to the middle cerebral circulation[9,10,14], since definitive clinical data in the microvasculature are still missing. The present model was then exploited to compare the cerebral hemodynamics during NSR and AF[9,10,14] at the same heart rate (75 bpm).

Similarly to the cardiovascular model, a network of compliances and resistances describes the cerebral circulation from the large arteries level up to the peripheral and capillary regions. The cerebral circulation is expressed in terms of pressure, volume, and flow rate, and can be divided into three principal regions: large arteries, distal arterial circulation, and capillary/venous circulation. The left vascular pathway ICA-MCA (i.e., internal carotid artery – middle cerebral artery) is here focused as representative of the blood flow and pressure distributions from large arteries to the capillary-venous circulation: left internal carotid artery ($P_a$ and $Q_{ICA,left}$), middle cerebral artery ($P_{MCA,left}$ and $Q_{MCA,left}$), middle distal district ($P_{dm,left}$ and $Q_{dm,left}$), and capillary-venous circulation ($P_c$ and $Q_{pv}$).

Details of the differential equations and model parameters are given elsewhere[9] and schematically recalled in the Supplementary Information. A representative sketch of the cerebral circulation is reported in Fig. 1c, together with examples of capillary pressures, $P_c$, in NSR and AF at different HRs.

**Data analysis.** First, the main statistics (mean, $\mu$, standard deviation, $\sigma$, coefficient of variation, cv) for the hemodynamic variables along the ICA-MCA pathway for 5000 cycles at different HRs during NSR and AF were computed. We also defined for each hemodynamic variable the damping factor, df, as the ratio between cv at lowest HR (i.e., 50 bpm) and cv at a certain HR. Regression analyses on HR-related variability of the damping factor, df, were performed, comparing AF and NSR regression slopes with ANCOVA (analysis of covariance) test.

Subsequently, the recurrence and distribution of critical events at different HRs was evaluated. We applied the definition of rare events during AF[9,10] to different HRs, so that each rare event during AF is defined in reference to the corresponding NSR at the corresponding HR. NSR outcomes were thus exploited to define the different reference thresholds for each HR, focusing on hypoperfusions and hypertensive events, the most meaningful events from the cerebral hemodynamic point of view. At a fixed HR, an hypoperfusion occurs when the average mean flow rate per beat stands below the threshold individuated by the 5$^{th}$ percentile in NSR at the corresponding HR. On the contrary, an hypertensive event takes place when the average pressure per beat is above the threshold individuated at the corresponding HR by the 95$^{th}$ percentile in NSR. Both hypoperfusions and hypertensive events can last one or more consecutive beats and this information is also retained. In Fig. 2, examples of hypoperfusions and hypertensive events are shown for HR = 50 and 130 bpm.

Setting 70 bpm as the reference simulation, Kolmogorov-Smirnov tests with 5% significance level were performed for AF simulation couples (70-50, 70–90, 70–110, 70–130 bpm) in order to test whether mean flow rate per beat and mean pressure per beat distributions differed in a statistically significant manner.

## Results
**Basic statistics.** Table 1 reports mean ($\mu$) and standard deviation ($\sigma$) values of the hemodynamic variables along the ICA-MCA pathway, namely $P_a$: systemic arterial pressure; $Q_{ICA,left}$: left internal carotid flow rate; $P_{MCA,left}$: left middle cerebral artery pressure; $Q_{MCA,left}$: left middle cerebral artery flow rate; $P_{dm,left}$: left middle distal pressure; $Q_{dm,left}$: left middle distal flow rate; $P_c$: cerebral capillary pressure; $Q_{pv}$: proximal venous flow rate.

Table 1 also shows coefficients of variation (cv): at the large arteries level ($Q_{ICA,left}$, $Q_{MCA,left}$, $P_a$, $P_{MCA,left}$), cv similarly decreases during NSR and AF at all HRs computed. In fact, at the corresponding HR, cv reported similar values during both NSR and AF, with relative variations not exceeding 10%. In addition, comparing HR-related variability damping factors for $Q_{MCA}$ and $P_a$, as depicted in Fig. 3a,b, no statistically significant difference of the regression lines emerged (p = 0.908 and p = 0.106 for flow rate and pressure, respectively).

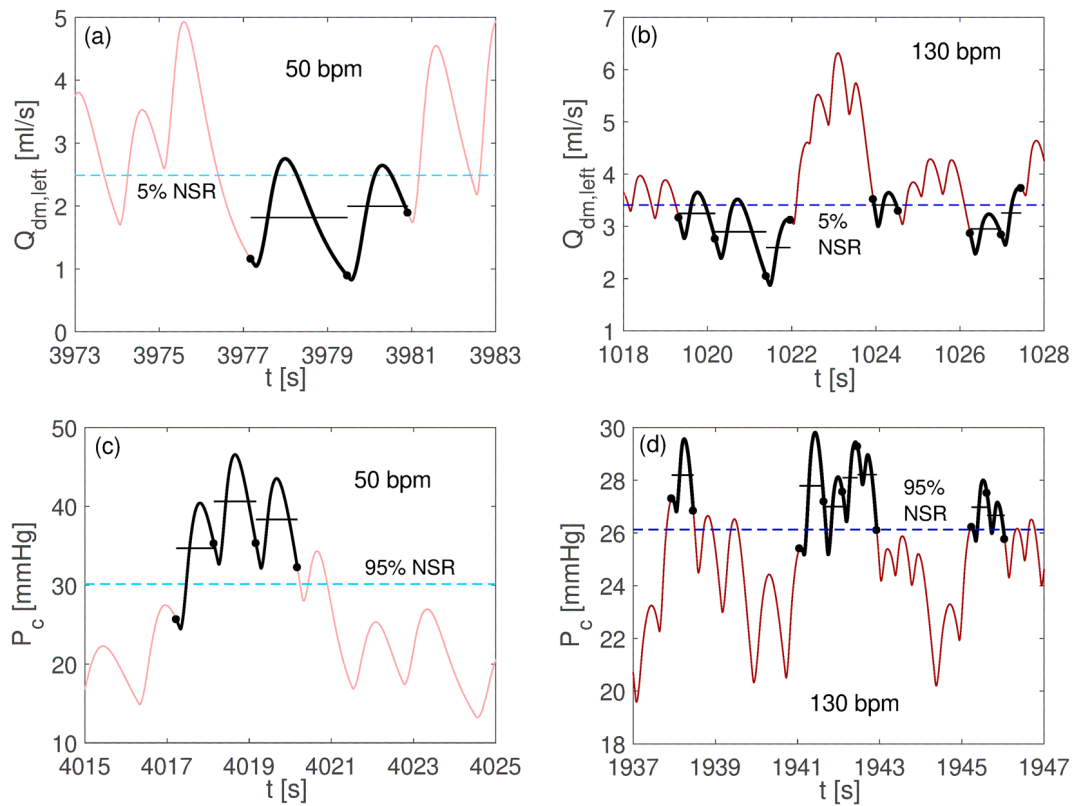

**Figure 2.** Examples of hypoperfusions and hypertensive events occurrence. Hypoperfusions for the distal flow rate in AF at 50 (panel a) and 130 (panel b) bpm, and hypertensive events for the capillary pressure in AF at 50 (panel c) and 130 (panel d) bpm. Average flow rate and pressure per beats are reported with black horizontal lines, the dashed blue horizontal lines represent the 5$^{th}$ (panels a and b) and 95$^{th}$ (panels c and d) percentile NSR thresholds, while black dots individuate the beginning/end of the RR beating.

Conversely, the hemodynamic situation appears significantly different at the distal-capillary level. At this site ($P_{dm,left}$, $P_c$, $Q_{dm,left}$, $Q_{pv}$), at lower HRs, AF presents cv values 1,3-1,5 times higher than NSR, while if the ventricular rate increases (from 50 to 130 bpm), AF variability becomes 1,7-2,5 times higher than NSR. If we compare HR-related variability damping factors for $Q_{dm,left}$ and $P_c$ (Fig. 3c,d), AF promotes a reduced ventricular-rate related damping of variability both concerning flow rate (p = 0.003) and pressure (p = 0.002), compared to NSR.

**Recurrence and distribution of rare events.** Table 2 reports the number of hypoperfusions and hypertensive events, stratified by AF simulations and HR, for all the hemodynamic variables of the ICA-MCA pathway. In Fig. 2 examples of hypoperfusion (the average mean flow rate per beat stands below the threshold individuated by the 5th percentile in NSR at the same corresponding HR) and hypertensive (the average pressure per beat is above the threshold individuated at the corresponding same HR by the 95th percentile in NSR) events are shown for HR = 50 (left panels) and 130 bpm (right panels). In general, at the systemic and proximal cerebral circle level, critical events were infrequent, being represented only at 50 bpm by 1 hypoperfusion at the carotid level and 2 hypoperfusions at middle cerebral artery level over 5000 beats.

In fact, hypoperfusions and hypertensive events mainly occurred in the distal circulation (downstream the middle cerebral artery), and progressively increased with HR (the only exception is $P_{dm,left}$, showing a "rapid grow-slow decrease" trend with HR). In particular, at 130 bpm AF simulations of the cerebral capillary level, 905 hypertensive and 534 hypoperfusions events over 5000 beats occurred, showing a two-threefold increase compared to AF simulations at lower HRs.

Figure 4 illustrates the absolute frequency (over 5000 cardiac cycles) of hypoperfusions and hypertensive events, respectively, as function of the consecutive beats involved for the distal-capillary variables ($Q_{dm,left}$, $P_c$). The Kolmogorov-Smirnov test (Supplementary Table 2 in the Supplementary Information) indicates that distributions of the computed hemodynamic variables significantly differ between each AF simulation (50, 90, 110, 130 bpm) and the reference 70 bpm simulation (all p values < 0.001).

|  | 50 bpm | 70 bpm | 90 bpm | 110 bpm | 130 bpm |
|---|---|---|---|---|---|
| **NSR** | | | | | |
| Pa [mmHg]: $\mu \pm \sigma$ | 90,09 ± 16,81 | 96,95 ± 14,36 | 100,95 ± 12,60 | 103,26 ± 11,25 | 104,24 ± 10,07 |
| cv | 0,19 | 0,15 | 0,12 | 0,11 | 0,10 |
| QICA, left [ml/s]: $\mu \pm \sigma$ | 4,69 ± 2,23 | 4,73 ± 1,82 | 4,75 ± 1,57 | 4,76 ± 1,41 | 4,77 ± 1,29 |
| cv | 0,48 | 0,38 | 0,33 | 0,30 | 0,27 |
| PMCA, left [mmHg]: $\mu \pm \sigma$ | 87,43 ± 15,59 | 94,25 ± 13,38 | 98,25 ± 11,78 | 100,55 ± 10,53 | 101,53 ± 9,43 |
| cv | 0,18 | 0,14 | 0,12 | 0,10 | 0,09 |
| QMCA, left [ml/s]: $\mu \pm \sigma$ | 3,70 ± 1,68 | 3,74 ± 1,35 | 3,75 ± 1,13 | 3,76 ± 0,99 | 3,77 ± 0,87 |
| cv | 0,45 | 0,36 | 0,30 | 0,26 | 0,23 |
| Pdm, left [mmHg]: $\mu \pm \sigma$ | 53,48 ± 4,85 | 56,93 ± 3,28 | 58,94 ± 2,32 | 60,09 ± 1,68 | 60,58 ± 1,28 |
| cv | 0,09 | 0,06 | 0,04 | 0,03 | 0,02 |
| Qdm, left [ml/s]: $\mu \pm \sigma$ | 3,70 ± 0,75 | 3,74 ± 0,50 | 3,75 ± 0,36 | 3,76 ± 0,26 | 3,77 ± 0,21 |
| cv | 0,20 | 0,13 | 0,10 | 0,07 | 0,06 |
| Pc [mmHg]: $\mu \pm \sigma$ | 24,88 ± 3,70 | 25,00 ± 2,28 | 25,04 ± 1,53 | 25,05 ± 1,07 | 25,05 ± 0,81 |
| cv | 0,15 | 0,09 | 0,06 | 0,04 | 0,03 |
| Qpv [ml/s]: $\mu \pm \sigma$ | 12,36 ± 2,81 | 12,46 ± 1,84 | 12,50 ± 1,28 | 12,53 ± 0,92 | 12,54 ± 0,70 |
| cv | 0,23 | 0,15 | 0,10 | 0,07 | 0,06 |
| **AF** | | | | | |
| Pa [mmHg]: $\mu \pm \sigma$ | 87,37 ± 18,04 | 94,20 ± 15,41 | 98,20 ± 13,21 | 100,42 ± 11,70 | 101,64 ± 10,48 |
| cv | 0,21 | 0,16 | 0,13 | 0,11 | 0,10 |
| QICA, left [ml/s]: $\mu \pm \sigma$ | 4,64 ± 2,32 | 4,71 ± 1,93 | 4,74 ± 1,65 | 4,75 ± 1,47 | 4,75 ± 1,35 |
| cv | 0,50 | 0,41 | 0,35 | 0,31 | 0,28 |
| PMCA, left [mmHg]: $\mu \pm \sigma$ | 84,73 ± 16,81 | 91,52 ± 14,44 | 95,51 ± 12,38 | 97,72 ± 10,98 | 98,94 ± 9,83 |
| cv | 0,20 | 0,16 | 0,13 | 0,11 | 0,10 |
| QMCA, left [ml/s]: $\mu \pm \sigma$ | 3,65 ± 1,79 | 3,72 ± 1,46 | 3,74 ± 1,21 | 3,75 ± 1,06 | 3,76 ± 0,94 |
| cv | 0,49 | 0,39 | 0,32 | 0,28 | 0,25 |
| Pdm, left [mmHg]: $\mu \pm \sigma$ | 52,06 ± 7,23 | 55,54 ± 5,68 | 57,56 ± 4,21 | 58,67 ± 3,37 | 59,28 ± 2,73 |
| cv | 0,14 | 0,10 | 0,07 | 0,06 | 0,05 |
| Qdm, left [ml/s]: $\mu \pm \sigma$ | 3,65 ± 1,05 | 3,72 ± 0,81 | 3,74 ± 0,60 | 3,75 ± 0,48 | 3,76 ± 0,40 |
| cv | 0,29 | 0,22 | 0,16 | 0,13 | 0,11 |
| Pc [mmHg]: $\mu \pm \sigma$ | 24,69 ± 5,45 | 24,94 ± 3,79 | 25,01 ± 2,64 | 25,03 ± 2,06 | 25,04 ± 1,68 |
| cv | 0,22 | 0,15 | 0,10 | 0,08 | 0,07 |
| Qpv [ml/s]: $\mu \pm \sigma$ | 12,22 ± 3,57 | 12,41 ± 2,62 | 12,47 ± 1,90 | 12,50 ± 1,50 | 12,51 ± 1,24 |
| cv | 0,29 | 0,21 | 0,15 | 0,12 | 0,10 |

**Table 1.** Mean ($\mu$), standard deviation ($\sigma$) and coefficient of variation (cv) of the hemodynamic variables along the selected ICA-MCA path, during NSR and AF at different HRs. $P_a$: systemic arterial pressure; $Q_{ICA,left}$: left internal carotid flow rate; $P_{MCA,left}$: left middle cerebral artery pressure; $Q_{MCA,left}$: left middle cerebral artery flow rate; $P_{dm,left}$: left middle distal pressure; $Q_{dm,left}$: left middle distal flow rate; $P_c$: cerebral capillary pressure; $Q_{pv}$: proximal venous flow rate.

## Discussion

Several hypotheses have been proposed in the attempt of explaining the, to date unknown, correlation between AF and cognitive impairment/dementia, in patients not presenting clinical cerebrovascular events (ictus/TIA)[5,28]. In particular, subclinical micro-embolic events, manifesting as silent cerebral ischemia (SCI) lesions at cerebral magnetic resonance imaging (MRI) scan[29,30], and cerebral microbleeds[31], possibly related to suboptimal oral anticoagulation therapy (OAT)[32–34], have candidate as potential contributors to this phenomenon. Another hypothesis, however, is that AF, by RR interval variability and loss of atrial systole, may produce cerebral hemodynamic alteration, possibly leading to reduced cerebral flow, brain damage and atrophy[35,36]. Moreover, our group previously demonstrated through a computational model of the cerebral circulation that AF, *per se*, is associated with transient and repetitive critical hemodynamic events in the deep circle (hypoperfusion and hypertensive events)[9,10], that could possibly relate to the genesis of a quote of non-microembolic SCIs and non-OAT related microbleeds.

Previous studies have explored the deleterious hemodynamic effect of an irregular heart rhythm, such as in AF, on systemic cardiovascular function, both in clinical[37–39] and experimental[40] settings. Irregular RR intervals were associated with a reduced cardiac output, increased pulmonary capillary wedge pressure and increased right atrial pressure, mainly due to beat-to-beat changes in ventricular filling, with short RR intervals decreasing cardiac output more than long RR intervals. Interestingly, Herbert[39] found that the correlation between the percentage of short cycle lengths and a decrease in cardiac index was more marked for patients with an average ventricular rate >75 beats/min than for those with a slower average rate. First evidence was provided that slow ventricular response in AF can partly compensate for hemodynamic consequences of an irregular ventricular rhythm, since at low heart rates even the shortest RR intervals may be sufficiently long to allow for adequate diastolic filling of the left ventricle. Of note, long term clinical follow-up studies apparently demonstrated a reverse association, indicating that a reduced RR variability/irregularity could be associated with higher mortality in AF patients[41–43]. These studies, however, were performed in AF patients with concomitant heart failure or severe valvular heart disease, thus the reduced irregularity of RR interval was likely a biomarker of underlying autonomic dysfunction, which directly reflects the presence of a functional/structural substrate associated with increased mortality. In addition, medical therapy recommended for heart failure and/or valvular heart disease may likely impact cardiovascular mortality independent of the achieved RR variability or irregularity.

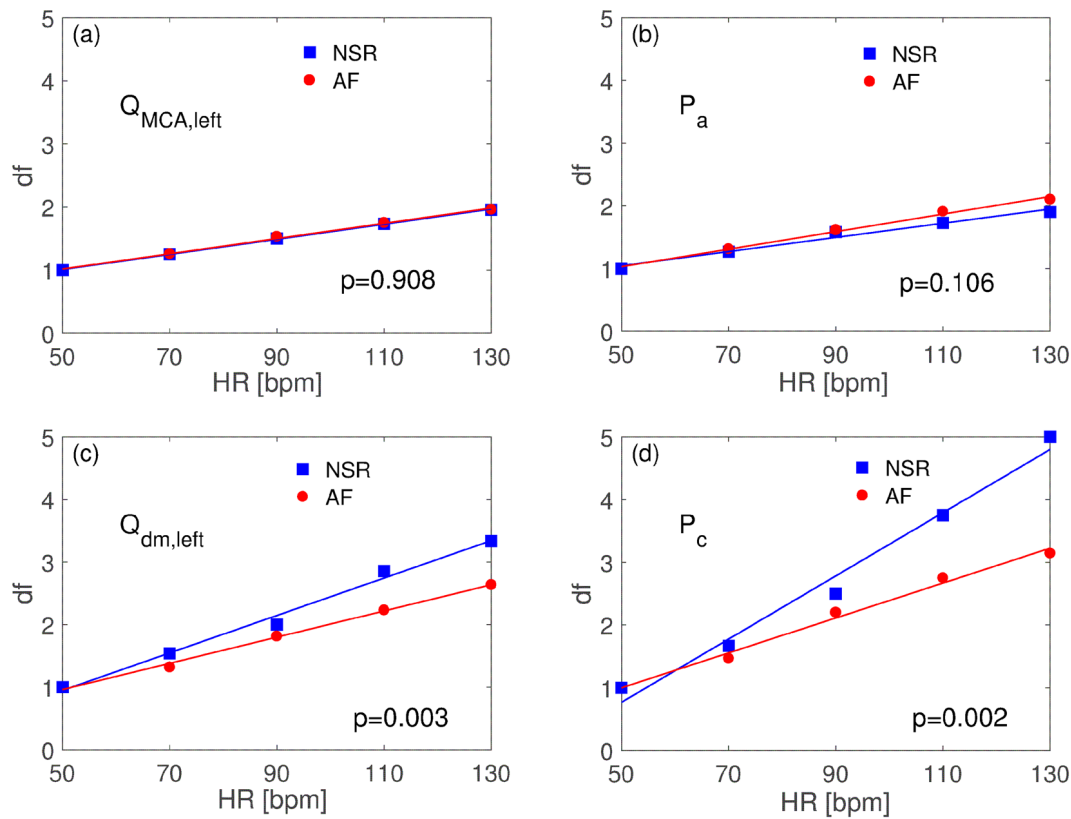

**Figure 3.** Variability damping factors, df, regression analysis. (**a**) $Q_{MCA,left}$ rate; (**b**) $P_a$; (**c**) $Q_{dm,left}$; (**d**) $P_c$. Reported ANCOVA p values test the null hypothesis that regression lines has the same slopes.

|  | $Q_{ICA,left}$ | $Q_{MCA,left}$ | $Q_{dm,left}$ | $Q_{pv}$ |
|---|---|---|---|---|
| **Hypoperfusions** | | | | |
| 50 bpm | 1 | 2 | 196 | 124 |
| 70 bpm | 0 | 0 | 321 | 136 |
| 90 bpm | 0 | 0 | 386 | 216 |
| 110 bpm | 0 | 0 | 451 | 352 |
| 130 bpm | 0 | 0 | 534 | 415 |
| **Hypertensive events** | | | | |
|  | $P_a$ | $P_{MCA,left}$ | $P_{dm,left}$ | $P_c$ |
| 50 bpm | 0 | 0 | 231 | 456 |
| 70 bpm | 0 | 0 | 478 | 549 |
| 90 bpm | 0 | 0 | 408 | 559 |
| 110 bpm | 0 | 0 | 354 | 811 |
| 130 bpm | 0 | 0 | 285 | 905 |

**Table 2.** Total number of rare one-beat events as function of the HR. (top) hypoperfusions, (bottom) hypertensive events.

In the present study, increasing ventricle response during AF was associated to an overall variability reduction of the computed cerebral hemodynamic variables, both in AF and in NSR, due to the signal-flattening effect induced by the shorter duration of the mean heart beat intervals. Moreover, AF presented higher variability in the signals, if compared to NSR, only in the distal cerebral circle (downstream the MCA, thus at the arteriolar-capillary level), underlining how the baroreceptor and autoregulation mechanisms (both implemented in the present modeling approach) are able to normalize AF-induced hemodynamic perturbations only at the systemic and proximal cerebral circle level. Still, variability of the signals decreased harsher in NSR than in AF at the distal cerebral circle level, differently from the systemic-proximal cerebral circle level where the dampening effect was similar between AF and NSR. This registered effect may, therefore, indicate that, at higher HR during AF, distal cerebral circle progressively loses the ability to absorb AF-induced hemodynamic perturbations, resulting in a limited HR-related dampening of signals variability if compared to NSR.

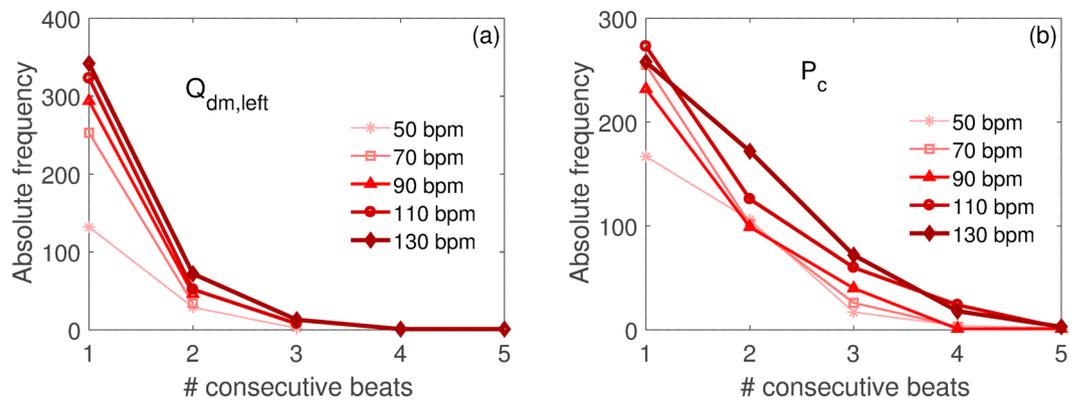

**Figure 4.** Absolute frequency over 5000 heartbeats of hypoperfusions and hypertensive events during AF in the distal circle (downstream MCA). (**a**) $Q_{dm,left}$, (**b**) $P_c$. The abscissa indicates the number of consecutive beats.

The reduced HR-related dampening of the variability of cerebral hemodynamic parameters in AF consequently leads to more frequent critical events (in terms of hypoperfusions and hypertensive events) at higher HRs. In this perspective, two aspects clearly emerge:

- Critical episodes mainly occur in the distal cerebral circle and significantly increase in frequency with HR. Since the trend is substantially monotone with HR, this suggests that there is no optimal HR target to minimize the occurrence of hypoperfusions and hypertensive episodes at this level of the cerebral circulation;
- In the 50 bpm-AF simulation (the lowest simulated HR in the present study), there is the potential occurrence (3 hypoperfusions events registered over 5000 beats) of critical events at the systemic-proximal cerebral circle level, while no critical events emerge for HR $\geq$70 bpm.

Given these findings, the optimal HR to minimize deep cerebral critical events by avoiding proximal cerebral hypoperfusions - in particular in patients with reduced baroreceptor mechanisms[44] - is above 50 and below 70 bpm. In fact, these results provide theoretical support to the clinical findings of Cacciatore *et al.*[13], reporting that low (<50 bpm) and high (>90 bpm) HR in AF patients are associated with worse cognitive outcomes.

To date, available clinical trials have failed to demonstrate a superiority of rhythm control and strict rate control in terms of classical mid-term hard cardiovascular outcomes[11,45]. However, these computational data, together with the results of Cacciatore *et al.*[13], suggest that in permanent AF a strict rate control strategy targeting resting HR around 70 bpm may be beneficial in terms of cerebral hemodynamics, since it minimizes deep critical events without increasing the risk of proximal cerebral hypoperfusions, possibly slowing the progressive onset of cognitive dysfunction in these patients. Based on the relevant impact in terms of quality of life and social costs of cognitive impairment and dementia, further clinical studies should necessarily focus on cerebral outcomes of NSR maintenance and ventricle rate response during permanent AF.

**Limitations.** The present computational model does not consider the impact that rate control drugs (e.g. digoxin, beta blockers, non-dihydropiridine calcium channel blockers) could exert on the cardiovascular system. Second, the cerebrovascular model assumes a perfectly functioning baroreceptor and cerebral autoregulation mechanism in both NSR and AF.

### Conclusions
In the present computational study ventricular response during AF exerts an impact on cerebral hemodynamics. In particular, higher ventricular rates relate to a progressive increase in critical cerebral events (hypoperfusions and hypertensive events) at the distal level (downstream the MCA). These results possibly suggest that, while all efforts should be addressed to maintain NSR as long as possible, a strict rate control strategy could be beneficial in terms of cognitive outcomes in patients with permanent AF.

### References

1. Zoni-Berisso, M., Lercari, F., Carazza, T. & Domenicucci, S. Epidemiology of atrial fibrillation: European perspective. *Clinical epidemiology* **6**, 213–220 (2014).
2. Graves, K. G. *et al.* Atrial fibrillation incrementally increases dementia risk across all CHADS2 and CHA2DS2VASc strata in patients receiving long-term warfarin. *American heart journal* **188**, 93–98 (2017).

3. Kalantarian, S., Stern, T. A., Mansour, M. & Ruskin, J. N. Cognitive impairment associated with atrial fibrillation: a meta-analysis. *Annals of internal medicine* **158**(5 Pt 1), 338–346 (2013).
4. Chen, L. Y. *et al.* Association of Atrial Fibrillation With Cognitive Decline and Dementia Over 20 Years: The ARIC-NCS (Atherosclerosis Risk in Communities Neurocognitive Study). *Journal of the American Heart Association* **7**(6), e007301 (2018).
5. Hui, D. S., Morley, J. E., Mikolajczak, P. C. & Lee, R. Atrial fibrillation: A major risk factor for cognitive decline. *American heart journal* **169**(4), 448–456 (2015).
6. Jacobs, V., Cutler, M. J., Day, J. D. & Bunch, T. J. Atrial fibrillation and dementia. *Trends in cardiovascular medicine* **25**(1), 44–51 (2015).
7. Rivard, L. & Khairy, P. Mechanisms, Clinical Significance, and Prevention of Cognitive Impairment in Patients With Atrial Fibrillation. *Canadian Journal of Cardiology* **33**(12), 1556–1564 (2017).
8. Purkayastha, S. & Sorond, F. Transcranial Doppler ultrasound: technique and application. *Semin Neurol* **32**(4), 411–420 (2012).
9. Anselmino, M., Scarsoglio, S., Saglietto, A., Gaita, F. & Ridolfi, L. Transient cerebral hypoperfusion and hypertensive events during atrial fibrillation: a plausible mechanism for cognitive impairment. *Sci Rep* **6**, 28635 (2016).
10. Scarsoglio, S., Saglietto, A., Anselmino, M., Gaita, F. & Ridolfi, L. Alteration of cerebrovascular haemodynamic patterns due to atrial fibrillation: an in silico investigation. *Journal of The Royal Society Interface* **14**(129), 20170180 (2017).
11. Van Gelder, I. C. *et al.* Lenient versus strict rate control in patients with atrial fibrillation. *The New England journal of medicine* **362**(15), 1363–1373 (2010).
12. Wyse, D. G. Lenient versus strict rate control in atrial fibrillation some devils in the details. *Journal of the American College of Cardiology* **58**(9), 950–952 (2011).
13. Cacciatore, F. *et al.* Role of ventricular rate response on dementia in cognitively impaired elderly subjects with atrial fibrillation: a 10-year study. *Dementia and geriatric cognitive disorders* **34**(3-4), 143–148 (2012).
14. Scarsoglio, S., Cazzato, F. & Ridolfi, L. From time-series to complex networks: Application to the cerebrovascular flow patterns in atrial fibrillation. *Chaos: An Interdisciplinary Journal of Nonlinear Science* **27**(9), 093107 (2017).
15. Hayano, J. *et al.* Spectral characteristics of ventricular response to atrial fibrillation. *The American journal of physiology* **273**(6 Pt 2), H2811–2816 (1997).
16. Scarsoglio, S., Guala, A., Camporeale, C. & Ridolfi, L. Impact of atrial fibrillation on the cardiovascular system through a lumped-parameter approach. *Medical & biological engineering & computing* **52**(11), 905–920 (2014).
17. Hennig, T., Maass, P., Hayano, J. & Heinrichs, S. Exponential distribution of long heart beat intervals during atrial fibrillation and their relevance for white noise behaviour in power spectrum. *Journal of biological physics* **32**(5), 383–392 (2006).
18. Goldberger, A. L. *et al.* PhysioBank, PhysioToolkit, and PhysioNet: components of a new research resource for complex physiologic signals. *Circulation* **101**(23), e215–e220 (2000).
19. Tateno, K. & Glass, L. Automatic detection of atrial fibrillation using the coefficient of variation and density histograms of RR and deltaRR intervals. *Medical & biological engineering & computing* **39**(6), 664–671 (2001).
20. Anselmino, M., Scarsoglio, S., Saglietto, A., Gaita, F. & Ridolfi, L. A computational study on the relation between resting heart rate and atrial fibrillation hemodynamics under exercise. *PloS one* **12**(1), e0169967 (2017).
21. Sosnowski, M., Korzeniowska, B., Macfarlane, P. W. & Tendera, M. Relationship between RR interval variation and left ventricular function in sinus rhythm and atrial fibrillation as estimated by means of heart rate variability fraction. *Cardiology journal* **18**(5), 538–545 (2011).
22. Korakianitis, T. & Shi, Y. Numerical simulation of cardiovascular dynamics with healthy and diseased heart valves. *Journal of biomechanics* **39**(11), 1964–1982 (2006).
23. Scarsoglio, S., Camporeale, C., Guala, A. & Ridolfi, L. Fluid dynamics of heart valves during atrial fibrillation: a lumped parameter-based approach. *Comput Methods Biomech Biomed Engin* **19**(10), 1060–1068 (2016).
24. Anselmino, M. *et al.* Rate Control Management of Atrial Fibrillation: May a Mathematical Model Suggest an Ideal Heart Rate? *PloS one* **10**(3), e0119868 (2015).
25. Scarsoglio, S., Saglietto, A., Gaita, F., Ridolfi, L. & Anselmino, M. Computational fluid dynamics modelling of left valvular heart diseases during atrial fibrillation. *PeerJ* **4**, e2240 (2016).
26. Ottesen J. T., Olufsen M. S., Larsen J. K. Applied Mathematical Models in Human Physiology: Society for Industial and Applied Mathematics; 2004.
27. Ursino, M. & Giannessi, M. A model of cerebrovascular reactivity including the circle of willis and cortical anastomoses. *Annals of biomedical engineering* **38**(3), 955–974 (2010).
28. Ihara, M. & Washida, K. Linking Atrial Fibrillation with Alzheimer's Disease: Epidemiological, Pathological, and Mechanistic Evidence. *Journal of Alzheimer's Disease* **62**(1), 61–72 (2018).
29. Gaita, F. *et al.* Prevalence of silent cerebral ischemia in paroxysmal and persistent atrial fibrillation and correlation with cognitive function. *Journal of the American College of Cardiology* **62**(21), 1990–1997 (2013).
30. Chen, L. Y. *et al.* Atrial fibrillation and cognitive decline-the role of subclinical cerebral infarcts: the atherosclerosis risk in communities study. *Stroke; a journal of cerebral circulation* **45**(9), 2568–2574 (2014).
31. Selim, M. & Diener, H.-C. Atrial fibrillation and microbleeds. *Stroke; a journal of cerebral circulation* **48**(10), 2660–2664 (2017).
32. Bunch, T. J. *et al.* Atrial Fibrillation Patients Treated With Long-Term Warfarin Anticoagulation Have Higher Rates of All Dementia Types Compared With Patients Receiving Long-Term Warfarin for Other Indications. *Journal of the American Heart Association* **5**(7), e003932 (2016).
33. Jacobs, V. *et al.* Time outside of therapeutic range in atrial fibrillation patients is associated with long-term risk of dementia. *Heart rhythm: the official journal of the Heart Rhythm Society* **11**(12), 2206–2213 (2014).
34. Jacobs V. *et al.* Percent INR Time With a Supratherapeutic INR in Atrial Fibrillation Patients Also Using an Antiplatelet Agent Is Associated With Long-Term Risk of Dementia. *Journal of cardiovascular electrophysiology* 2015.
35. Stefansdottir, H. *et al.* Atrial fibrillation is associated with reduced brain volume and cognitive function independent of cerebral infarcts. *Stroke; a journal of cerebral circulation* **44**(4), 1020–1025 (2013).
36. Gardarsdottir M. *et al.* Atrial fibrillation is associated with decreased total cerebral blood flow and brain perfusion. *EP Europace* (2017).
37. Daoud, E. G. *et al.* Effect of an irregular ventricular rhythm on cardiac output. *American Journal of Cardiology* **78**(12), 1433–6 (1996).
38. Clark, D. M., Plumb, V. J., Epstein, A. E. & Kay, G. N. Hemodynamic effects of an irregular sequence of ventricular cycle lengths during atrial fibrillation. *Journal of the American College of Cardiology* **30**(4), 1039–45 (1997).
39. Herbert, W. H. Cardiac output and the varying R-R interval of atrial fibrillation. *Journal of Electrocardiology* **6**(2), 131–5 (1973).
40. Naito, M., David, D., Michelson, E. L., Schaffenburg, M. & Dreifus, L. S. The hemodynamic consequences of cardiac arrhythmias: evaluation of the relative roles of abnormal atrioventricular sequencing, irregularity of ventricular rhythm and atrial fibrillation in a canine model. *American heart journal* **106**(2), 284–91 (1983).
41. Stein, K. M., Borer, J. S., Hochreiter, C., Devereux, R. B. & Kligfield, P. Variability of the ventricular response in atrial fibrillation and prognosis in chronic nonischemic mitral regurgitation. *American Journal of Cardiology* **74**(9), 906–11 (1994).
42. Frey, B. *et al.* Diurnal variation of ventricular response to atrial fibrillation in patients with advanced heart failure. *American heart journal* **129**(1), 58–65 (1995).

43. Cygankiewicz, I. *et al.* Reduced Irregularity of Ventricular Response During Atrial Fibrillation and Long-term Outcome in Patients With Heart Failure. *American Journal of Cardiology* **116**(7), 1071–5 (2015).
44. Monahan, K. D. Effect of aging on baroreflex function in humans. *American Journal of Physiology-Regulatory, Integrative and Comparative Physiology* **293**(1), R3–R12 (2007).
45. Van Gelder, I. C. *et al.* Rate Control Efficacy in permanent atrial fibrillation: a comparison between lenient versus strict rate control in patients with and without heart failure. Background, aims, and design of RACE II. *American heart journal* **152**(3), 420–426 (2006).

### Acknowledgements
This study was performed thanks to the support of the "Compagnia di San Paolo" within the project "Progetti di Ricerca di Ateneo – 2016: Cerebral hemodynamics during atrial fibrillation (CSTO 160444)" of the University of Turin, Italy. Funded author: Dr. M. Anselmino. The funders had no role in study design, data collection and analysis, decision to publish, or preparation of the manuscript.

### Author Contributions
All authors conceived and designed the experiments. S.S. performed the experiments. M.A., S.S., A.S. and L.R. analyzed the data. S.S. and L.R. contributed reagents/materials/analysis tools. All authors wrote and reviewed the manuscript.

### Additional Information
**Supplementary information** accompanies this paper at https://doi.org/10.1038/s41598-019-40445-5.

**Competing Interests:** The authors declare no competing interests.

**Publisher's note:** Springer Nature remains neutral with regard to jurisdictional claims in published maps and institutional affiliations.